\documentclass[usenatbib,usegraphicx]{mn2e}  
\usepackage{subfigure}  
\usepackage{longtable}  
\usepackage{graphicx}  
\usepackage{graphics}  
\usepackage{keyval}  
\usepackage{trig}  
  
\def\beq{\begin{equation}}  
\def\eeq{\end{equation}}  
\def\bey{\begin{eqnarray}}  
\def\eey{\end{eqnarray}}

\def\grad{{\bf \nabla}}  
  
\def\a0{$a_0$}

\def\lta{{_ <\atop{^\sim}}}  
  
\title[Escaping from MOND]{Escaping from Modified Newtonian Dynamics}  
\author[B. Famaey, J.-P. Bruneton and H.S. Zhao]{Benoit Famaey$^{1}$\thanks{FNRS Research Associate; email: bfamaey@ulb.ac.be}, Jean-Philippe Bruneton$^{2}$ and HongSheng Zhao$^{3}$\\  
$^{1}$Institut d'Astronomie et d'Astrophysique, Universit\'e Libre  
de Bruxelles,  
CP 226, Bvd du Triomphe, B-1050, Bruxelles, Belgium\\  
$^{2}$Institut d'Astrophysique de Paris, UMR 7095-CNRS, Universit\'e Pierre et Marie Curie-Paris 6, 98 bis Bvd Arago F-75014, Paris, France\\  
$^{3}$SUPA, School of Physics and Astronomy, University of St. Andrews, Fife KY16 9SS, UK\\}  
  
\begin{document}  
  
\date{Accepted ... Received ... ; in original form ...}  
  
\pagerange{\pageref{firstpage}--\pageref{lastpage}} \pubyear{2006}  
  
\maketitle  
  
\label{firstpage}  
  
\begin{abstract}  
We present a new test of modified Newtonian dynamics (MOND) on  
galactic scales, based on the escape speed in the Solar  
Neighbourhood. This test is independent from other empirical  
successes of MOND at reproducing the phenomenology of galactic  
rotation curves. The galactic escape speed in MOND is entirely  
determined by the baryonic content of the Galaxy and the external  
field in which it is embedded. We estimate that the external field  
in which the Milky Way must be embedded to produce the observed  
local escape speed of $\sim 550$~${\rm km} \, {\rm s}^{-1}$ is of the  
order of $a_0/100$, where $a_0$ is the dividing acceleration scale  
below which gravity is boosted in MOND. This is compatible with the external gravitational field actually acting on the Milky Way.  
\end{abstract}  
  
\begin{keywords}  
Galaxy: kinematics and dynamics - Gravitation - Dark Matter  
\end{keywords}  
  
\section{Introduction}  
  
The speed needed by a star in the Solar Neighbourghood to escape  
the local gravitational field of the Milky Way -- known as the  
local galactic escape speed -- has long been known to be an  
essential indicator of the amount of mass beyond the  
galactocentric radius of the Sun (e.g. Binney \& Tremaine  
1987). Given the poor constraints on the shape of the galactic  
rotation curve beyond that radius (e.g. Binney \& Dehnen 1997;  
Famaey \& Binney 2005, herefater FB05), it has become one of the  
most convincing arguments in favour of a halo of dark matter in  
our Galaxy (Smith et al. 2007). However, the observed tight  
correlation between the mass profiles of baryonic matter and dark  
matter at all radii in spiral galaxies (e.g. McGaugh et al. 2007; Famaey et al. 2007) rather lends support to modified Newtonian dynamics (MOND). This theory postulates that for accelerations  
below $a_0 = 1.2 \times 10^{-10} {\rm m} \, {\rm s}^{-2}$ the effective  
gravitational attraction approaches $(g_N a_0)^{1/2}$ where $g_N$  
is the usual Newtonian gravitational field. Without resorting to galactic dark matter, this simple prescription is amazingly successful at reproducing many aspects of the dynamics of galaxies that were unknown when Milgrom (1983) introduced the theory (e.g. Sanders \& McGaugh 2002; Tiret \& Combes 2007).  It is however worth noting that the theory might still welcome (and might even need) hot dark matter, which could condense in galaxy clusters and boost the (modified) gravitational potential on Mpc scale (Sanders 2003; Angus et al. 2007). 

In MOND, the gravitational potential $\Phi$ of an isolated point
mass $M$ at a distance $r \gg (GM/a_0)^{1/2}$ is of the form: \beq
\label{eqn:potentielMondien}\Phi(r) \sim (GMa_0)^{1/2} \, {\rm
ln}(r). \eeq This implies that the associated rotation curve is
asymptotically flat, but in turn it implies that there is no
escape from such an isolated point mass. In practice however, no
objects are truly isolated in the Universe and this has wider and
more subtle implications in MOND than in Newton-Einstein gravity,
for the very reason that MOND breaks the Strong Equivalence
Principle (SEP). In particular, this letter illustrates how the external field regularizes the above divergent potential, so that it \textit{is} possible to escape from non-isolated point masses in MOND.

More generally, this breakdown of the SEP remarkably allows to
derive properties of the gravitational field in which a system is
embedded from its internal dynamics (and not only from tides).
For this reason, we are able to compute hereafter the external
gravitational field in which the Milky Way should be embedded in
MOND in order to account for the measured local galactic escape
speed. The result is in good agreement with the gravitational
field that Large Scale Structures are expected to exert on the
Galaxy (of the order of $a_0/100$). The MOND paradigm thus
successfully explains the value of the local escape speed without
referring to any halo of dark matter.
    
\section{Observed galactic escape speed}  
  
The local galactic escape speed in the Solar Neighbourhood can be  
estimated by using local samples of high velocity stars (Fich \&  
Tremaine 1991), but this task was until recently hampered by the  
very small number of such observed stars and by the fact that they  
were selected from proper-motion surveys. Now, radial velocity  
data from the Radial Velocity Experiment (RAVE), a  
magnitude-limited survey avoiding kinematical biases (Steinmetz et  
al. 2006), have been used to constrain the local escape speed  
(Smith et al. 2007). A total of 33 stars (16 from the RAVE survey  
and 17 from Beers et al. 2000) with velocities higher than  
300~${\rm km} \, {\rm s}^{-1}$ have been selected, and, following  
the technique of Leonard \& Tremaine (1990), the local escape  
speed has been estimated to lie in the following 90\% confidence  
interval: \beq \protect\label{eqn:esc} 498 \, {\rm km} \, {\rm  
s}^{-1} < v_{\rm esc} < 608 \, {\rm km} \, {\rm s}^{-1}, \eeq with  
a median likelihood of 544~${\rm km} \, {\rm s}^{-1}$. Any unbound star in the sample could have biased the  result but Yu \& Tremaine (2003) have shown that hypervelocity stars ejected from the galactic center are actually very rare (i.e. 1 every $10^5$ years) and thus not likely to belong to such a survey.

\section{Galactic escape speed in MOND}  
  
\subsection{The external field effect}  
  
The local galactic escape speed essentially depends on the profile  
of the gravitational potential in the outskirts of the Milky Way. The latter may of course depend on the precise realization of the MOND  
phenomenology. Here we assume that the gravitational potential  
$\Phi$ of the Milky Way satisfies the modified Poisson equation (Bekenstein \& Milgrom 1984): \beq \protect\label{eqn:poissoneffective}  
\nabla.[\mu(|-\grad\Phi + {\bf g}_e|/a_0) \nabla \Phi ] \simeq 4 \pi G  
\rho, \eeq where $\Phi$ is the non-external part of the potential, $\rho$ is the density of the (baryonic) matter,  
${\bf g}_e$ is the approximatively constant external field in  
which the Milky Way is embedded, and $\mu(x)$ is a function which  
runs from $\mu(x)=1$ at $x\gg 1$ to $\mu(x)=x$ at $x\ll 1$, and which can thus never be smaller than $|{\bf g}_e|/a_0$. Note however that Eq.~(\ref{eqn:poissoneffective}) is only an approximate and effective way to take into account the effect of the external field on local physics, in order to avoid solving the modified Poisson equation of Bekenstein \& Milgrom (1984) with an external source term $\rho_{\rm ext}$ on the right-hand side. This type of problem can be solved analytically only in some special configurations. For example, if $|\grad\Phi| \ll |{\bf g}_e|$ everywhere in the system, a full perturbative analytical solution can be found (Milgrom 1986; Zhao \& Tian 2006), but this is physically relevant only for the internal potential of low-density dwarf spheroidal galaxies orbiting more massive galaxies such as the Milky Way.

In the Milky Way, if $|{\bf g}_e| \ll a_0 \ll |\grad\Phi|$ (e.g. near the Galactic center), $\mu \approx 1$, and the whole dynamics is purely Newtonian. If $|{\bf g}_e| \ll |\grad\Phi| \ll a_0$ (e.g. at 50~kpc from the Galactic center), then the dynamics is in the MOND regime, but the external field can be neglected, leading to a potential given by Eq.~(\ref{eqn:potentielMondien}) up to a constant. Finally, when $|\grad\Phi| \ll |{\bf g}_e| \ll a_0$ in the outskirts of the system, $\mu$ tends to its asymptotic  
value $\mu(|{\bf g}_e|/a_0) \simeq |{\bf g}_e|/a_0$, and the  
gravitational potential is thus Newtonian with a renormalized  
gravitational constant. Then, contrary to the potential of Eq.~(\ref{eqn:potentielMondien}), it is possible to escape from such a potential.  This saturation of the $\mu$-function is the
essential piece needed to get a \textit{finite} but nevertheless
\textit{boosted} escape speed (compared to its Newtonian value).

This can be understood intuitively as follows. At the
phenomenological level, the MOND force deriving from
Eq.~(\ref{eqn:potentielMondien}) can equivalently be interpreted
in Newtonian gravity as providing an universal profile of dark
halos. Explicitly, the effective dark mass $M_{\rm eff}$ at radius
$r \gg (GM/a_0)^{1/2}$ reads $M_{\rm eff}(r) \sim M \times [(a_0)^{1/2}r
/(G M)^{1/2} -1]$ and thus diverges with $r$. Interpreted in
Newtonian gravity, MOND therefore predicts that any
\textit{isolated} baryonic mass $M$ creates an effective dark halo of
infinite mass with density profile of the form $\rho_{\rm eff} \propto
M_{\rm eff}(r)/r^3$. The external field effect however saturates the
$\mu$-function at large distance, thereby leading to a potential
which is logarithmic only on a finite range $(G M / a_0)^{1/2} \ll r \lta
r_e$, where $r_e$ is the transition radius at which the external
field equals the local one: $|{\bf g}_e|=|\grad \Phi|$. The external field thus truncates the effective dark halo at $r \approx r_e$, yielding a finite effective dark mass surrounding the Galaxy, which may explain the oberved value of the local escape speed (Eq.~\ref{eqn:esc}), depending on the actual value of ${\bf g}_e$.

\subsection{The escape speed} 

To simplify the problem, we consider herafter the spherically symmetric case.  The escape speed $v_{\rm esc}$ at radius $r$ is defined in a spherical potential as \beq \frac{1}{2}v_{\rm esc}^2(r) =  \Phi(\infty) - \Phi(r). \eeq We assume that the spatial variation of the external field is so small compared to the size of the Milky Way that it is approximately constant over all space, so that $\Phi(\infty)$ is well-defined.
We use the  
assumptions and values from the best analytic MOND model of the Milky Way  
(Model III of FB05). This means that we assume that all the  
baryonic mass is confined to $r<R_0$, where $R_0=8$~kpc is the  
galactocentric radius of the Sun, and that the asymptotic circular velocity is $v_{\infty}=165$~${\rm km} \, {\rm s}^{-1}$. The total baryonic mass $M$ of the Milky Way is related to the latter by means of the Tully-Fisher relation $G M  
a_0 = v_{\infty}^4$. Finally the interpolating $\mu$-function is  
chosen to be given by $\mu(x)=x/(1+x)$. In addition to providing the best analytic fit to the terminal velocity curve of the Milky Way (FB05), this function has been shown to yield excellent fits of rotation curves in external galaxies (Famaey et al. 2007). 

We then consider as a toy-model a centripetal external field acting as a constant force directed towards the Galactic Center, in order to estimate the order of magnitude of the external field needed to produce the observed escape speed. We thus solve Eq.~(\ref{eqn:poissoneffective}) for $g=|\grad\Phi|$ (we also note $g_e = |{\bf  
g}_e|$). Using spherical symmetry and hence Gauss theorem, this  
means solving: \beq \protect\label{eqn:g} \frac{a_0 \, g\,  
(g+g_e)}{a_0+g+g_e}=\frac{v_{\infty}^4}{r^2}. \eeq  
  
Then we compute the local escape speed 
\beq
v_{\rm esc}(R_0) = \left( 2\int_{R_0}^\infty g(r) {\rm d}r \right)^{1/2},
\eeq
and find that in order to have a local escape speed in the range allowed by Eq.~(\ref{eqn:esc}), with a  
median of 544~${\rm km} \, {\rm s}^{-1}$, the external field must be \beq  
\protect\label{ext1} g_e=0.010_{-0.008}^{+0.015} \times a_0 . \eeq  Note that, in addition to the intrinsic non-sphericity of the inner Milky Way itself, the real external field points in a definite direction, meaning that the escape speed contours are in reality non-spherical, depending on the angle between the internal and external force vectors. As a first order test of robustness, note that replacing $|{\bf g}+{\bf g}_e|=(g+g_e)$ by, e.g., $(g^2+g_e^2)^{1/2}$ in Eq.~(\ref{eqn:g}) only leads to a 20~${\rm km} \, {\rm s}^{-1}$ difference in the computed local escape speed for an external field $g_e$ of order $a_0/100$.
  
The Milky Way potential (whose zero is fixed at $R_0$) obtained from Eq.~(\ref{eqn:g}) is plotted on Figure 1. Overplotted are the corresponding MOND potential without external field (asymptoting to Eq.~\ref{eqn:potentielMondien}), and the Keplerian potential with  
renormalized gravitational constant $Ga_0/g_e$. The actual  
potential interpolates smoothly between these two behaviours, the  
transition occurring when the internal and external acceleration  
are of the same order of magnitude.

\begin{figure}  
\includegraphics[angle=0,width=8.5cm]{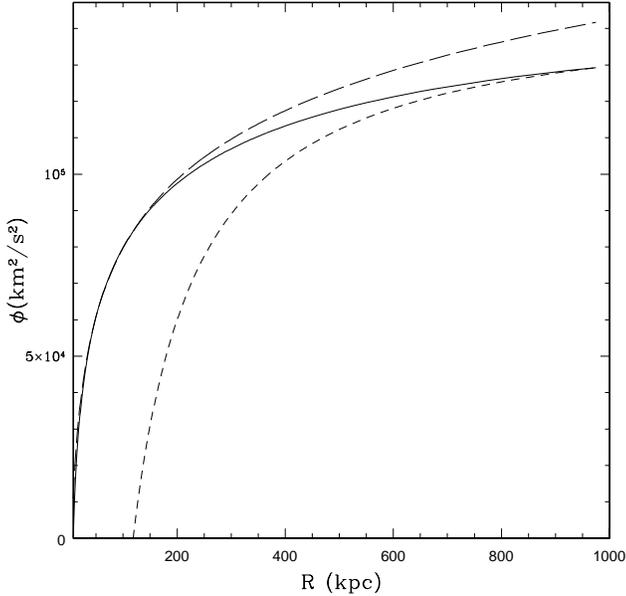}  
\caption{Full line: the MOND potential of the Milky Way for the  
parameters of FB05 $R_0=8$~kpc and  
$v_\infty=165$~${\rm km} \, {\rm s}^{-1}$, and an external field directed towards the Galactic center $g_e=0.01  \times a_0$. The potential is such that $\Phi(R_0)=0$, and it  
asymptotes to $\Phi(\infty) = 1/2 v_{\rm esc}^2(R_0)$ with $v_{\rm  
esc}(R_0)=544$~${\rm km} \, {\rm s}^{-1}$. Long-dashed line: the corresponding MOND potential  
without the external field. Short-dashed line: the Keplerian potential  
with a renormalized gravitational constant to which the MOND  
potential asymptotes.}\label{fig:potential}  
\end{figure}  
  
\section{Estimating the external field}  
  
In this section we focus on the various known sources that may 
produce the needed external field (Eq.~\ref{ext1}) in a MOND Universe. One must however keep in mind that when there is more than one attractor the MOND gravity is not simply the sum of the contributions.

There are two obvious possible contributions: the Andromeda 
galaxy M31, and the Large Scale Structure (LSS). Andromeda (e.g. Evans et al. 2000) is located at a distance $d_A \approx 800$~kpc and has a baryonic mass $M_A \approx 10^{11} \, M_\odot$. This means that the gravitational force per unit mass exerted by M31 on the Milky Way is of the order of
\beq
\frac{(GM_Aa_0)^{1/2}}{d_A} \simeq a_0/100,
\eeq
with some variations depending on the actual location considered in the Milky Way. However, since both galaxies have a roughly similar baryonic mass, the MOND gravitational field created by M31 is always smaller than the one created by the Milky Way in the direction opposite to M31, meaning that M31 cannot be the only source of the approximately constant external field we are looking for.

On the other hand, estimating the total contribution from all the relevant distant sources in the LSS combined with the one of M31 is an extremely arduous task (the MOND gravity not being the sum of the individual contributions), and would require a detailed knowledge of the precise geometry and mass content of the MOND Universe at large. However, one can estimate that a gravitational field of order $a_0/100$ is roughly the MOND acceleration 30~Mpc away from a galaxy cluster with an asymptotic, isotropic, line-of-sight velocity dispersion of 500~${\rm km} \, {\rm s}^{-1}$ (or 130~Mpc away from a cluster with a 1000~${\rm km} \, {\rm s}^{-1}$ dispersion). This means that the Virgo and Coma clusters could both individually contribute external fields of this magnitude (Milgrom 2002). 

Moreover, it has been known for twenty years (Lynden-Bell et al. 1988) 
that the Local Group belongs to a 600~${\rm km} \, {\rm 
s}^{-1}$ outflow w.r.t. the reference frame of 
the Cosmic Microwave Background. This is presumably due to the 
local gravitational attraction of LSS, and mainly from the 
so-called Great Attractor region at $(l \approx 325^\circ, b \approx -7^\circ$) (see Radburn-Smith et al. 2006). 
Once again, precisely calculating the value of this gravitational attraction at the location of the Milky Way in a MOND Universe is virtually impossible, but one can estimate its order of magnitude from the acceleration endured by the Local Group during a Hubble time in order to attain 
a peculiar velocity of 600~${\rm km} \, {\rm s}^{-1}$: \beq H_0 
\times 600 \, {\rm km} \, {\rm s}^{-1} \simeq a_0/100, \eeq for 
$H_0 = 70 \, {\rm km} \, {\rm s}^{-1} {\rm Mpc}^{-1}$. This estimation is independent from the assumption that we are living (or not) in a MOND Universe. In conclusion, M31, the Virgo and Coma clusters, as well as the LSS at large, {\it all} provide comparable gravitational fields that precisely lie in the allowed range (Eq.~\ref{ext1}) in order to produce the right local galactic escape speed. 
  
\section{Discussion}  

Using the best analytic MOND fit for the terminal velocity curve of the Milky Way (Model III of FB05), and considering a toy external field acting as a constant force directed towards the Galactic Center, we have estimated the external field in which the Milky Way must be embedded in a MOND Universe in order to produce the observed local escape speed (Sect.~2, Eq.~\ref{eqn:esc}). We found that the needed external field is of the order of $a_0/100$ (Sect.~3, Eq.~\ref{ext1}). 

One might question how this result depends on the specific
choice for the interpolating function $\mu$. As we explained at
the end of Sect.~3.1, the escape speed is mostly
determined by the value of the transition radius $r_e$ where the
internal and external accelerations are of the same order of
magnitude. The fact that the needed external field is well below $a_0$ means that the internal field at $r_e$ is in the deep-MOND regime $\mu(x)=x$  for any realistic $\mu$-function. However the $\mu$-function plays a non-trivial role in determining the stellar mass-to-light ratio needed to fit the terminal velocity curve of the Milky Way. For instance, using the ``standard" function $\mu(x)=x/(1+x^2)^{1/2}$ requires a higher mass-to-light ratio, leading to an asymptotic velocity $v_{\infty} = 175$~${\rm km} \, {\rm s}^{-1}$ (Model I of FB05). We calculate that it typically
leads to a 30~${\rm km} \, {\rm s}^{-1}$ increase in the local
escape speed for an external field of order $a_0/100$. Such a
variation is thus small compared to the error bars quoted in
Eq.~(\ref{eqn:esc}), and this signals the robustness of our
estimate w.r.t. the choice of $\mu$. On the other hand, our Eq.~(\ref{eqn:g}) is only an approximation for a couple of reasons: it neglects the presence of the distant external source(s) (and hence the direction of the external field), as well as a term proportional to the gradient of $\mu$ in Eq.~(\ref{eqn:poissoneffective}). However, both effects are proportional to $g_e$, meaning that our result does hold qualitatively as a first order estimate.

We argue that $a_0/100$ is a value roughly compatible with the gravitational attraction that Large Scale Structures are expected to exert on the Milky Way (Sect.~4). We thus conclude that the local galactic escape speed can be explained within a modified gravity framework \`a la MOND without the need of huge quantities of dark mass on galactic scales. However, it is very plausible that taking into account the matter density from M31, the Virgo and Coma clusters as well as all the structures within the Great Attractor in a MOND potential solver (Ciotti, Londrillo \& Nipoti 2006) could perhaps lead to a local gravitational field slightly above the range allowed by Eq.~(\ref{ext1}) (note that an important issue regarding this calculation might be the possibility of a scale-dependence of $a_0$; see e.g. the values used in Nusser 2002 and Skordis et al. 2006). Such an increase of the external field would slightly decrease the corresponding galactic escape speed, and might signal the need for small quantities of dark mass (in baryonic or non-baryonic form) at large radii from the Galactic center (e.g. in the form of 2~eV neutrinos; see Sanders 2003; Angus et al. 2007).

As a final note, we stress that there are good theoretical reasons to impose the $\mu$-function to lie within the range $\varepsilon<\mu(x)<1$ for all $x$, 
where $\varepsilon$ is a small dimensionless parameter. Indeed, if 
the $\mu$-function asymptotes in the standard way at $x \to 0$ (as 
$\mu(x) \approx x$) then the energy of the gravitational field 
surrounding a massive body is infinite, as was already pointed out 
in Bekenstein \& Milgrom (1984). Moreover, the most successful 
relativistic theory of the MOND paradigm to date (Bekenstein 2004), may exhibit dynamical singularities if the free function is chosen in such a way that the MOND phenomenology is recovered with an effective $\mu$-function asymptoting to zero at $x \to 0$ (Bruneton 2007). These two singularities are manifestly avoided if one instead uses a slight deviation from Milgrom's first proposal, namely $\mu(x) \sim \varepsilon + x$ (Sanders 1986; Famaey et al. 2007). This $\varepsilon$-term acts precisely as the toy external field that we have used in Sect.~3. Comparison 
with observed rotation curves, and especially from the last 
observed radius of NGC~3741 (Gentile et al. 2007) were used in Famaey et al. (2007) to constrain $\varepsilon$ to be lower than $1/10$. 
The present analysis of the local escape speed actually enables to put a 
stronger bound on this parameter. If $\varepsilon$ were of the order 
of $1/10$, it would saturate the $\mu$-function too early 
and the predicted escape speed would be significantly lower than 
the one observed. As a conclusion, the value of the local galactic escape speed constrains $\varepsilon$ to be $\varepsilon \lta 1/100$. 

\section*{acknowledgements}
We are grateful to Garry Angus, Alain Jorissen, Gianfranco Gentile, Pavel Kroupa, Holger Baumgardt, Bob Sanders and Adi Nusser for insightful comments on the manuscript.

\bsp  
  
\label{lastpage}

\end{document}